\documentclass{ws-ijmpcs}

\usepackage{array}
\usepackage{slashed}

\def\trimmarks{}

\DeclareMathOperator{\tr}{Tr}

\newcommand{\st}{{\scriptscriptstyle T}}
\newcommand{\sL}{{\scriptscriptstyle L}}

\begin{document}

\markboth{M. G. A. Buffing and P. J. Mulders
}
{Universality of Quark and Gluon TMD Correlators}

%%%%%%%%%%%%%%%%%%%%% Publisher's Area please ignore %%%%%%%%%%%%%%%
%
\catchline{}{}{}{}{}
%
%%%%%%%%%%%%%%%%%%%%%%%%%%%%%%%%%%%%%%%%%%%%%%%%%%%%%%%%%%%%%%%%%%%%

\title{UNIVERSALITY OF QUARK AND GLUON TMD CORRELATORS}

\author{M. G. A. BUFFING}

\address{Nikhef and Department of Physics and Astronomy, VU University Amsterdam\\
De Boelelaan 1081, NL-1081 HV Amsterdam, the Netherlands\\
m.g.a.buffing@vu.nl}

\author{P. J. MULDERS}

\address{Nikhef and Department of Physics and Astronomy, VU University Amsterdam\\
De Boelelaan 1081, NL-1081 HV Amsterdam, the Netherlands\\
mulders@few.vu.nl}

\author{A. MUKHERJEE}

\address{Department of Physics, Indian Institute of Technology Bombay\\
Mumbai 400076, India\\
asmita@phy.iitb.ac.in}

\maketitle
\vspace{10mm}
%\begin{history}
%\received{Day Month Year}
%\revised{Day Month Year}
%\end{history}

%%%%%%%%%%%%%%%%
\begin{abstract}
%%%%%%%%%%%%%%%%

Transverse Momentum Dependent (TMD) parton distribution functions (PDFs), in short referred to as TMDs, also take into account the transverse momentum ($p_\st$) of the partons. Just as the $p_\st$-integrated analogues we want to link them to quark and gluon matrix elements using Operator Product Expansion methods in QCD, involving operators of definite twist. The TMDs also involve operators of higher twist, which are not suppressed by powers of the hard scale, however. Using the expression for TMDs involving nonlocal matrix elements of quark and gluon fields there is a gauge link dependence, which also introduces an inherent process dependence. Using transverse moments, which are specific $p_\st$-weightings, we can establish the link with quark and gluon fields including the higher twist ones. We introduce (a finite number of) universal TMDs of definite rank and show how the process dependent TMDs can be written as combinations of these universal functions.
\end{abstract}
\ccode{PACS numbers: 12.38.-t, 13.85.Ni, 13.85.Qk}

%%%%%%%%%%%%%%%%%%%%%%%%%%%%%%%%%%%%%%%%%%%%%%%%%%%%%%%%%%%%%%%%%%%%%%%%%%%%%%%
\section{Introduction}
%%%%%%%%%%%%%%%%%%%%%%%%%%%%%%%%%%%%%%%%%%%%%%%%%%%%%%%%%%%%%%%%%%%%%%%%%%%%%%%

In high-energy processes parton distribution functions (PDF) and parton fragmentation functions (PFF) are used to describe the link between hadrons in initial and final states, respectively, as well as the quark and gluons that are used to describe the processes within the framework of Quantum Chromodynamics (QCD). The starting points are forward matrix elements of parton fields, such as the quark-quark correlator
\begin{equation}
\Phi_{ij}(p|p) =\int \frac{d^4\xi}{(2\pi)^4}\ e^{i\,p\cdot \xi}\ \langle P\vert \overline\psi_j(0)\,\psi_i(\xi)\vert P\rangle ,
\label{e:phi-basic}
\end{equation}
where a summation over color indices is understood, i.e. the operator structure $\tr_c\left( \psi_i(\xi)\overline\psi_j(0)\right)$ is used. This correlator replaces the fermionic correlator $\Phi \propto (\rlap{/}p + m)$ that one is used to for a single incoming fermion. Similarly one has the gluon correlator
\begin{equation}
\Gamma^{\mu\nu}(p|p) ={\int}\frac{d^4\xi}{(2\pi)^4}\ e^{ip\cdot\xi}\ \langle P\vert F^{n\mu}(0)\,F^{n\nu}(\xi)\vert P\rangle,
\label{e:gluonbasic}
\end{equation}
again with a color summation understood, i.e. $\tr_c\left( F^{n\mu}(0)\,F^{n\nu}(\xi)\right)$ is used as the operator structure if one uses for the gluon fields matrices in color triple representation. Besides a single quark, also a quark and a gluon from a particular hadron taking part in the hard process need to be accounted for. This is included as a multi-parton correlator, such as the quark-quark-gluon correlator
\begin{equation}
\Phi^\mu_{A\,ij}(p-p_1,p_1|p) = 
\int \frac{d^4\xi\,d^4\eta}{(2\pi)^8}
\ e^{i\,(p-p_1)\cdot \xi}\ e^{i\,p_1\cdot \eta}
\ \langle P\vert \overline\psi_j(0)\,A^\mu(\eta)\,\psi_i(\xi)\vert P\rangle.
\label{e:quarkgluonquark}
\end{equation}
The basic idea is to factorize these hadronic correlators (soft parts) in a full diagrammatic approach and parametrize them in terms of PDFs. This requires high energies in which case the momenta of different hadrons obey $P{\cdot} P^\prime \propto Q^2$, where $Q^2$ is the hard scale in the process. The hadronic momenta can be treated as light-like vectors $P$ and the hard process brings in a conjugate light-like vector $n$ such that $P{\cdot} n = 1$, for instance $n = P^\prime/P{\cdot} P^\prime$. With the light-like vectors one makes a Sudakov expansion of the parton momenta,
\begin{equation}
p = x P + p_\st + (p{\cdot} P - x M^2)n,
\end{equation}
with $x = p^+ = p{\cdot} n$. In any contraction with vectors outside the correlator, the component $xP$ contributes at order $Q$, the transverse component at order $M \sim Q^0$ and the remaining component contributes at order $M^2/Q$. This allows consecutive integration of the components to obtain from the fully unintegrated correlator in Eq.~(\ref{e:phi-basic}) the TMD light-front (LF) correlator
\begin{equation}
\Phi_{ij}(x,p_\st;n) =
\left. \int \frac{d\,\xi{\cdot} P\,d^2\xi_\st}{(2\pi)^3}\ e^{i\,p\cdot \xi}
\ \langle P\vert \overline\psi_j(0)\,\psi_i(\xi)\vert P\rangle
\right|_{\xi{\cdot} n = 0} ,
\label{e:phi-TMD}
\end{equation}
the collinear light-cone (LC) correlator
\begin{equation}
\Phi_{ij}(x) =
\left. \int \frac{d\,\xi{\cdot} P}{2\pi}\ e^{i\,p\cdot \xi}
\ \langle P\vert \overline\psi_j(0)\,\psi_i(\xi)\vert P\rangle
\right|_{\xi{\cdot} n = \xi_\st = 0\ \mbox{or}\ \xi^2 = 0} ,
\label{e:phi-x}
\end{equation}
or the local matrix element
\begin{equation}
\Phi_{ij} =
\left. \langle P\vert \overline\psi_j(0)\,\psi_i(\xi)\vert P\rangle
\right|_{\xi = 0} .
\end{equation}
The importance of integrating at least the light-cone (minus) component $p^- = p{\cdot} P$ is that the expression is at equal time, i.e.\ time-ordering is not relevant anymore for TMD or collinear PDFs.\cite{Diehl:1998sm} For local matrix elements one can calculate the anomalous dimensions, which show up as the Mellin moments of the splitting functions that govern the scaling behavior of the collinear correlator $\Phi(x)$. We note that the collinear correlator is not simply an integrated TMD. The dependence on an upper limit $\Phi(x;Q^2) = \int^{Q}d^2p_\st\ \Phi(x,p_\st)$ is found from the anomalous dimensions (splitting functions). One has an $\alpha_s/p_\st^2$ behavior of TMDs that is calculable using collinear TMDs and which matches to the intrinsic nonperturbative $p_\st$-behavior.\cite{Collins:1984kg} We note that in operator product expansion language, the collinear correlators involve operators of definite twist, while TMD correlators involve operators of various twist.

%%%%%%%%%%%%%%%%%%%%%%%%%%%%%%%%%%%%%%%%%%%%%%%%%%%%%%%%%%%%%%%%%%%%%%%%%%%%%%%
\section{Color Gauge Invariance}
%%%%%%%%%%%%%%%%%%%%%%%%%%%%%%%%%%%%%%%%%%%%%%%%%%%%%%%%%%%%%%%%%%%%%%%%%%%%%%%

In order to determine the importance of a particular correlator in a hard process, one can do a dimensional analysis to find out when they contribute in an expansion in the inverse hard scale. Dominant are the ones with lowest canonical dimension obtained by maximizing contractions with $n$, for instance for quark or gluon fields the minimal canonical dimensions dim[$\overline\psi(0)\rlap{/}n\,\psi(\xi)$] = dim[$F^{n\alpha}(0)\,F^{n\beta}(\xi)$] = 2, while an example for a multi-parton combination gives dim[$\overline\psi(0)\rlap{/}n\,A_\st^\alpha(\eta)\,\psi(\xi)$] = 3. Equivalently, one can maximize the number of $P$'s in the parametrization of $\Phi_{ij}$. Of course one immediately sees that any number of collinear $n{\cdot} A(\eta) = A^n(\eta)$ fields doesn't matter. Furthermore one must take care of color gauge invariance, for instance when dealing with the gluon fields and one must include derivatives in color gauge-invariant combinations. With dimension zero there is $iD^n = i\partial^n + gA^n$ and with dimension one there is $iD_\st^\alpha = i\partial_\st^\alpha+gA_\st^\alpha$. The color gauge-invariant expressions for quark and gluon distribution functions actually include gauge link operators,
\begin{equation}
U_{[0,\xi]} = {\cal P}\exp\left(-i\int_0^{\xi} d\zeta_\mu A^\mu(\zeta)\right),
\end{equation}
connecting the nonlocal fields,
\begin{eqnarray}
&&\Phi_{ij}^{[U]}(x,p_\st;n) =
\int \frac{d\,\xi{\cdot} P\,d^2\xi_\st}{(2\pi)^3}\ e^{i\,p\cdot \xi}
\ \langle P\vert \overline\psi_j(0)\,U_{[0,\xi]}\,\psi_i(\xi)\vert P\rangle
\Biggr|_{\textrm{\scriptsize{LF}}} ,
\label{e:qmatrix}
\\
&&\Gamma^{[U,U^\prime]\,\mu\nu}(x,p_\st;n) =
{\int}\frac{d\,\xi{\cdot}P\,d^2\xi_\st}{(2\pi)^3}\ 
e^{ip\cdot\xi}
\nonumber \\ && \mbox{}\qquad\qquad\qquad \times
\tr\,\langle P{,}S|\,F^{n\mu}(0)\,
U_{[0,\xi]}^{\phantom{\prime}}\,
F^{n\nu}(\xi)\,U_{[\xi,0]}^\prime\,
|P{,}S\rangle\Biggr|_{\textrm{\scriptsize{LF}}} .
\label{e:matrix}
\end{eqnarray}
For transverse separations, the gauge links involve paths running along the minus direction to $\pm \infty$ (dimensionally preferred), which are closed with one or more transverse pieces at light-cone infinity. The two simplest possibilities are $U_{[0,\xi]}^{[\pm]}$ = $U^n_{[0,\pm\infty]}\,U^T_{[0_\st,\xi_\st]}\,U^n_{[\pm\infty,\xi]}$, leading to gauge link dependent quark TMDs $\Phi_q^{[\pm]}(x,p_\st)$.\cite{Belitsky:2002sm,Bomhof:2004aw} For gluons, the correlator involves color gauge-invariant traces of field-operators $F^{n\alpha}$, requiring the inclusion of \emph{two} gauge links $U_{[0,\xi]}$ and $U_{[\xi,0]}^\prime$. Again the simplest possibilities are the past- and future-pointing gauge links $U^{[\pm]}$, giving even in the simplest case four gluon TMDs $\Gamma^{[\pm,\pm^\dagger]}(x,p_\st)$. In general, many gauge link structures are possible, connecting the positions $0$ and $\xi$ in different ways. Furthermore, there are also contributions containing Wilson loops, given by $U^{[\square]}=U_{[0,\xi]}^{[+]}U_{[\xi,0]}^{[-]}$ = $U_{[0,\xi]}^{[+]}U_{[0,\xi]}^{[-]\dagger}$ or $U^{[\square]\dagger}$ = $U_{[0,\xi]}^{[-]}U_{[\xi,0]}^{[+]}$ = $U_{[0,\xi]}^{[-]}U_{[0,\xi]}^{[+]\dagger}$. A detailed list of useful gauge link structures can be found in Ref.~\refcite{Buffing:2013kca}.

\begin{figure}
\begin{center}
\epsfig{file=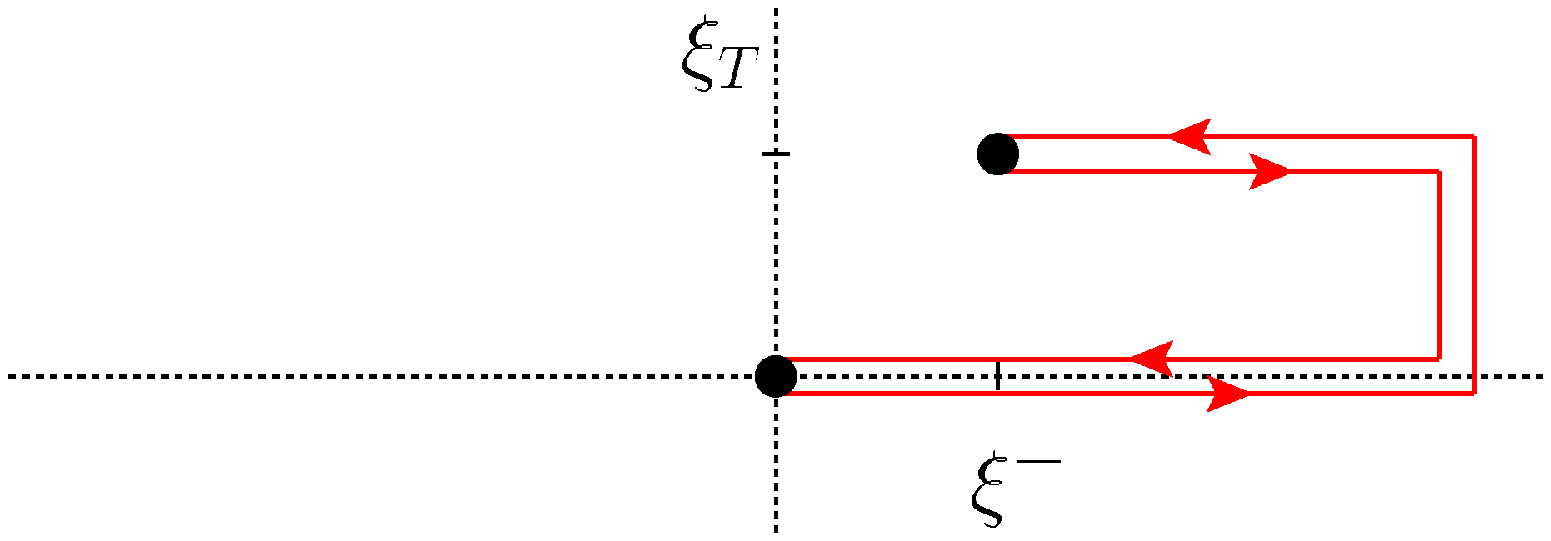,width=0.35\textwidth}
\hspace{15mm}
\epsfig{file=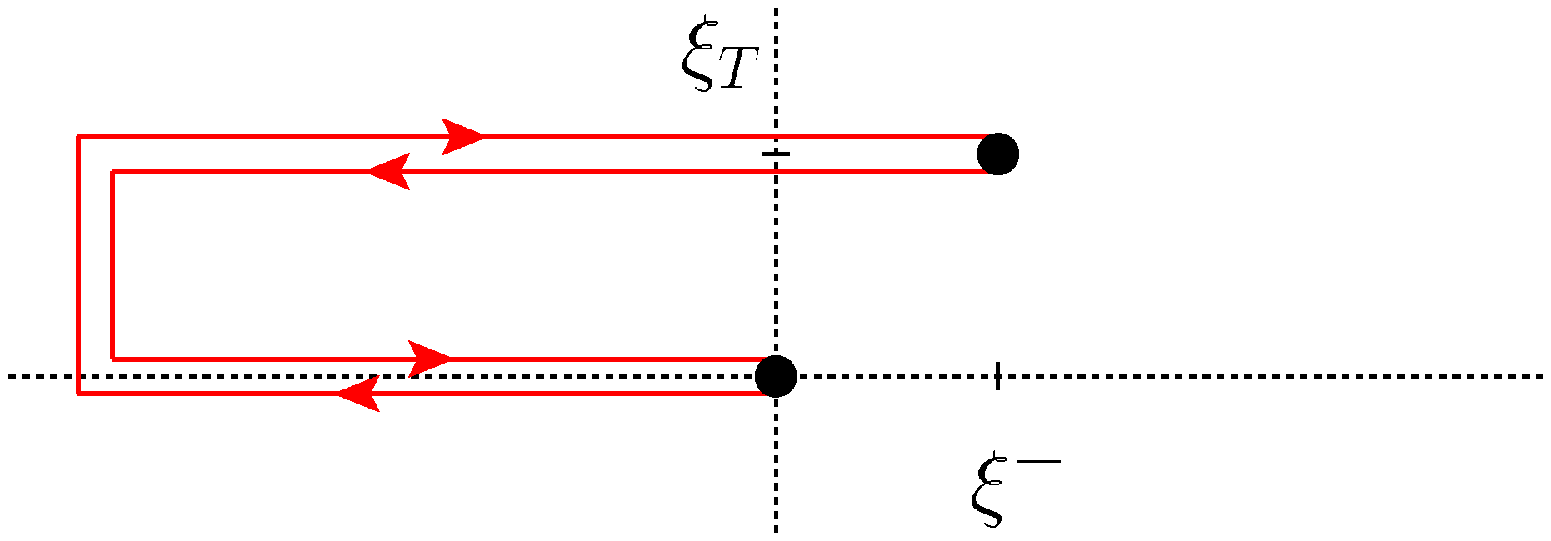,width=0.35\textwidth}
\\[1mm]
(a)\hspace{56mm} (b)
\\[3mm]
\epsfig{file=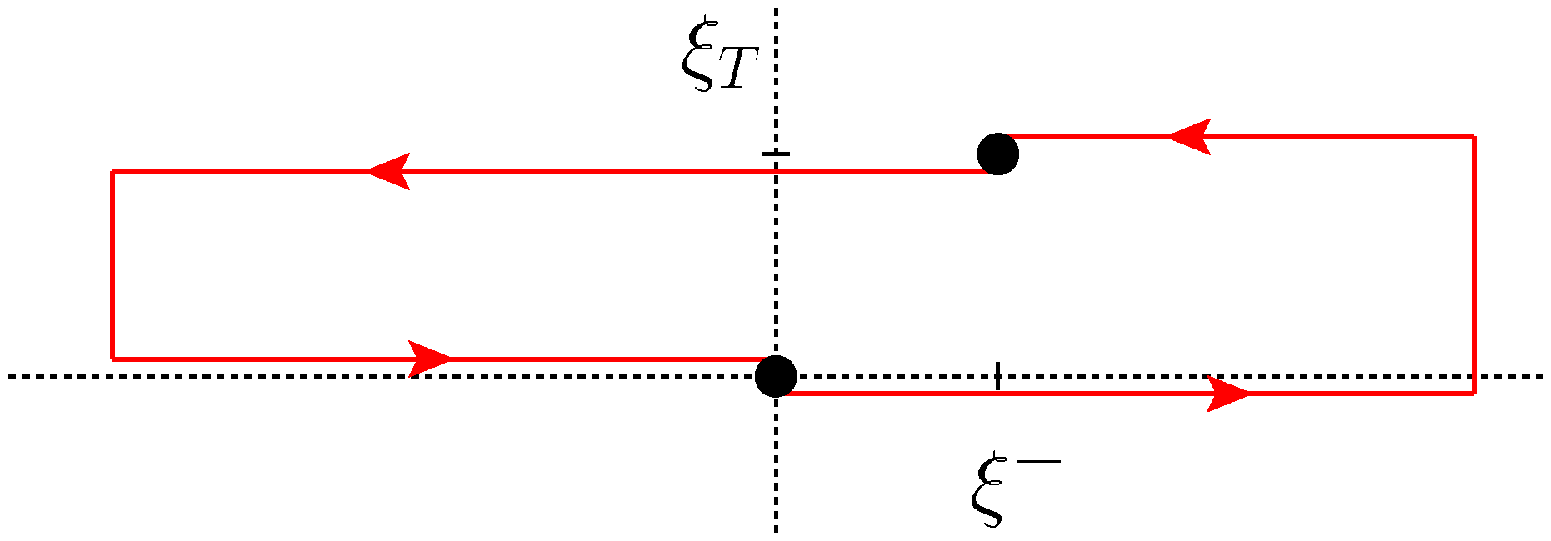,width=0.35\textwidth}
\hspace{15mm}
\epsfig{file=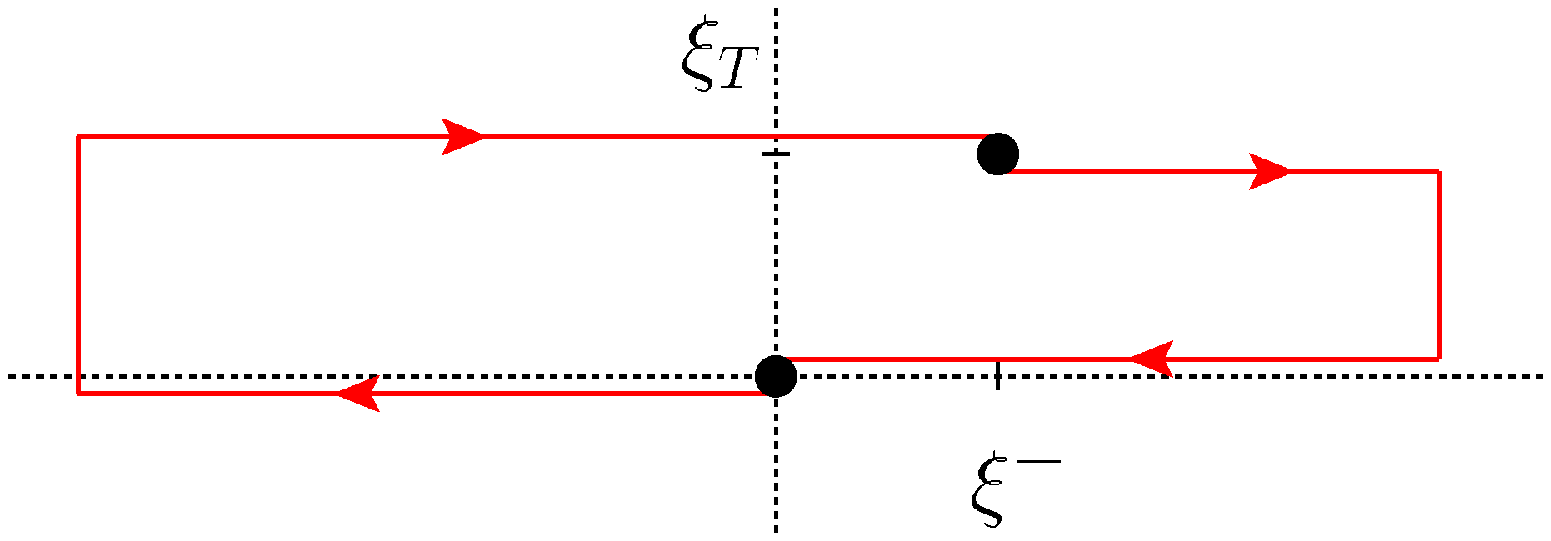,width=0.35\textwidth}
\\[1mm]
(c)\hspace{56mm} (d)
\\[3mm]
\epsfig{file=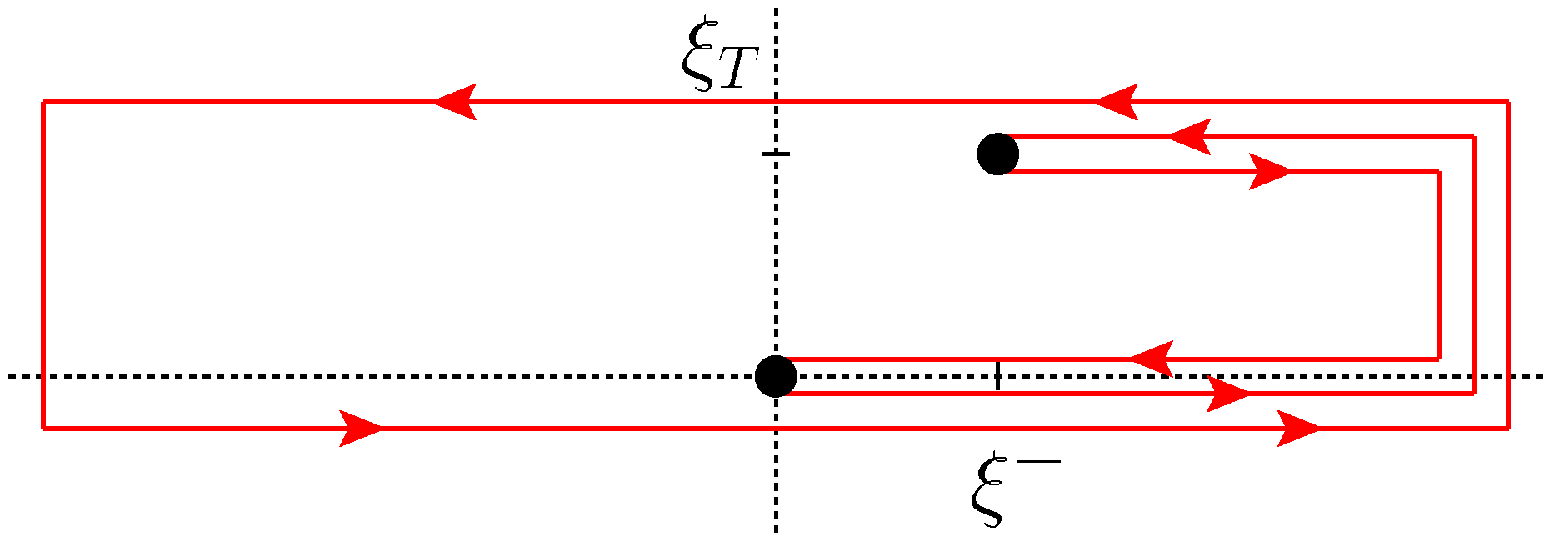,width=0.35\textwidth}
\hspace{15mm}
\epsfig{file=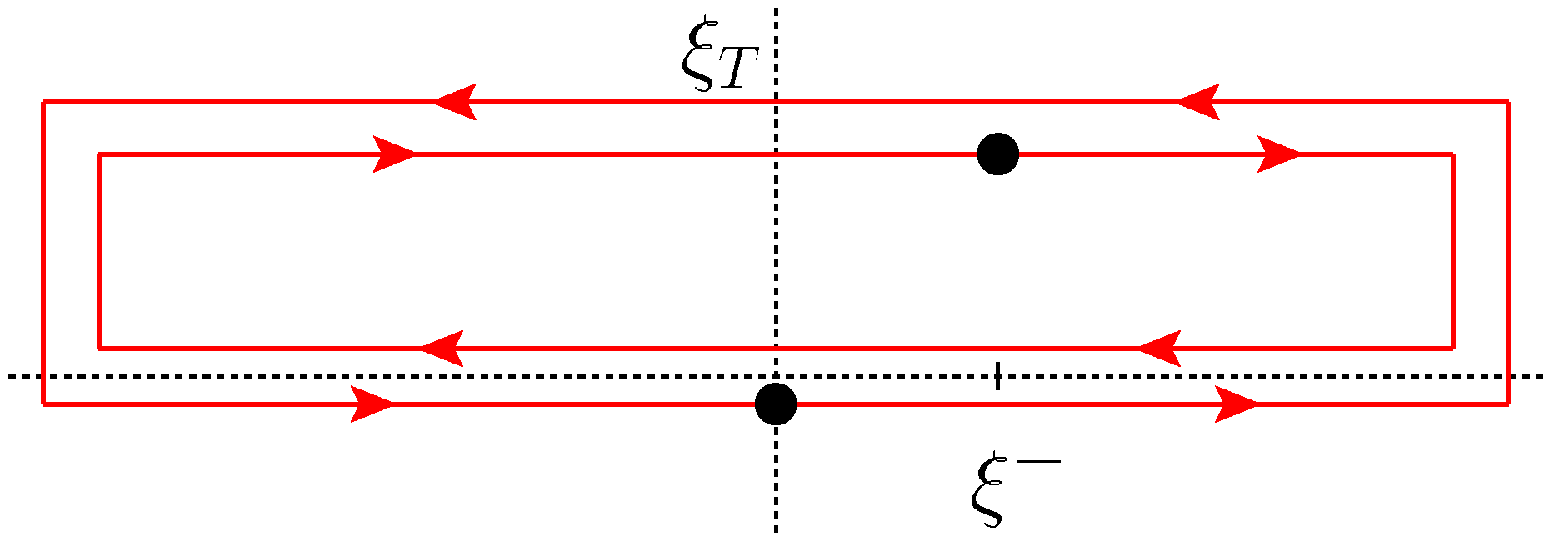,width=0.35\textwidth}
\\[1mm]
(e)\hspace{56mm} (f)
\caption{\label{f:GL}
Illustrations of a number of gauge link structures $[U,U^\prime]$. In these figures, the two big dots represent the coordinates of $0$ and $\xi$. The horizontal axis is the light-cone direction $n_-$ and the vertical axis represents the transverse directions. The four simplest gauge link structures are
(a) $[+,+^{\dagger}]$, 
(b) $[-,-^{\dagger}]$, 
(c) $[+,-^{\dagger}]$ and 
(d) $[-,+^{\dagger}]$.
An example of a type 2 gauge link structure is (e) $[+,+^{\dagger}(\square)]$, while (f) $[(F(0)\square),(F(\xi)\square^{\dagger})]$ is an example of a type 3 gauge link structure.}
\end{center}
\end{figure}

\begin{figure}
\begin{center}
\epsfig{file=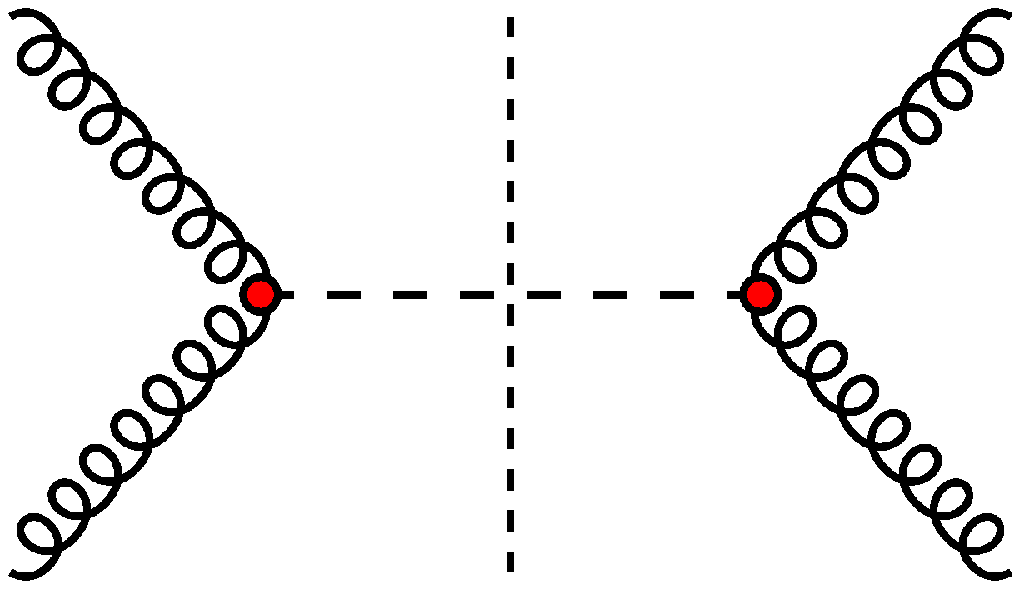,width=0.20\textwidth}
\hspace{5mm}
\epsfig{file=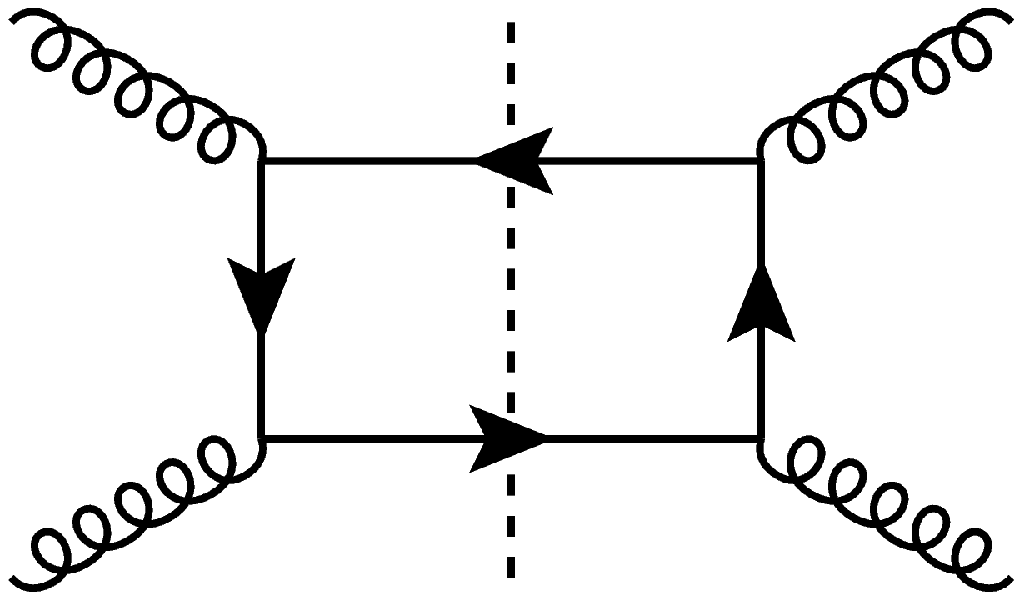,width=0.20\textwidth}
\hspace{5mm}
\epsfig{file=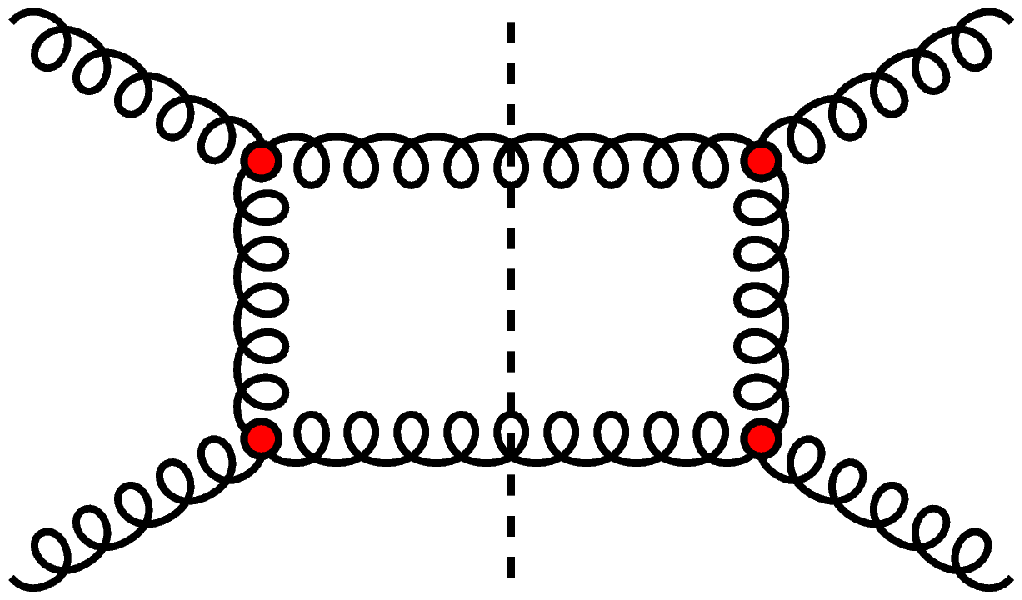,width=0.20\textwidth}
\hspace{5mm}
\epsfig{file=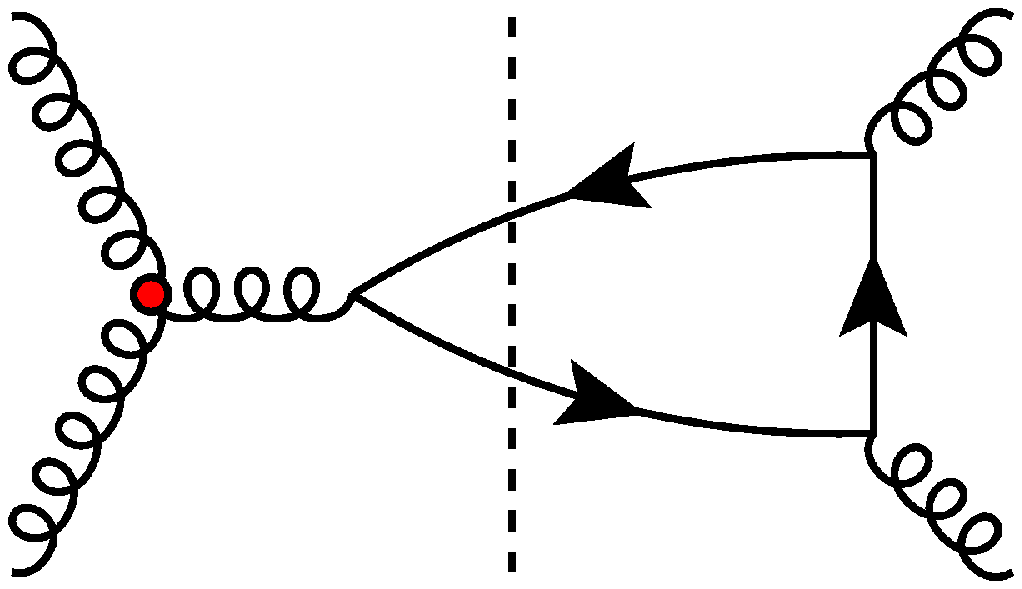,width=0.20\textwidth}
\\[1mm]
(a)\hspace{26mm} (b) \hspace{26mm} (c) \hspace{26mm} (d)
\caption{\label{f:FD}
Four hard scattering diagrams playing a role in proton-proton interactions, see main text for a description.}
\end{center}
\end{figure}

\begin{figure}
\begin{center}
\epsfig{file=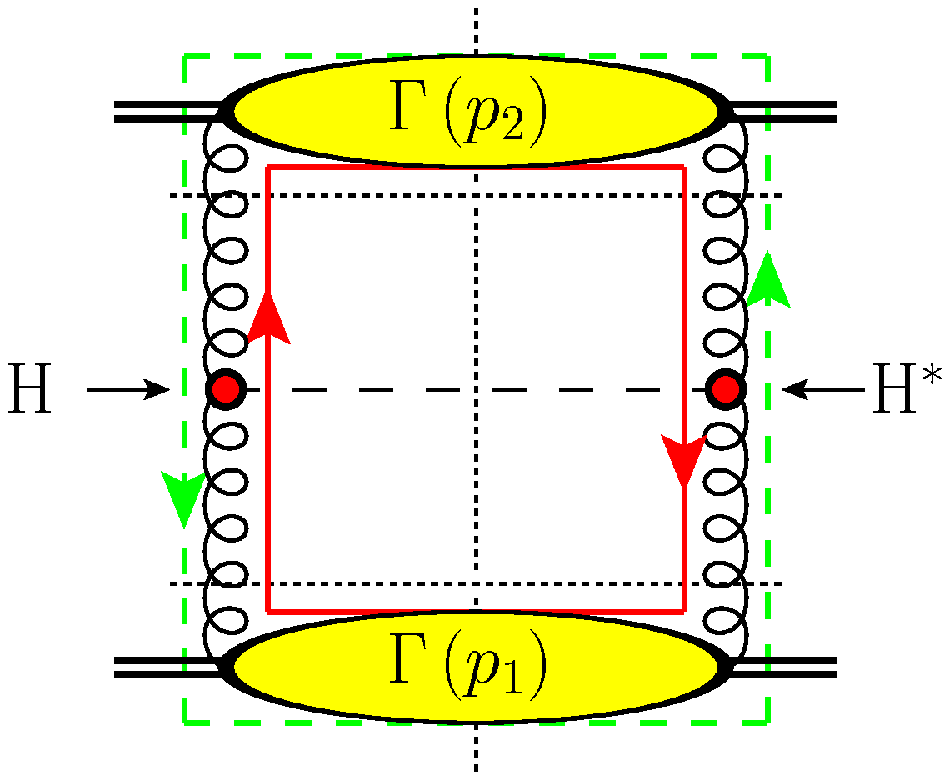,width=0.3\textwidth}
\hspace{15mm}
\epsfig{file=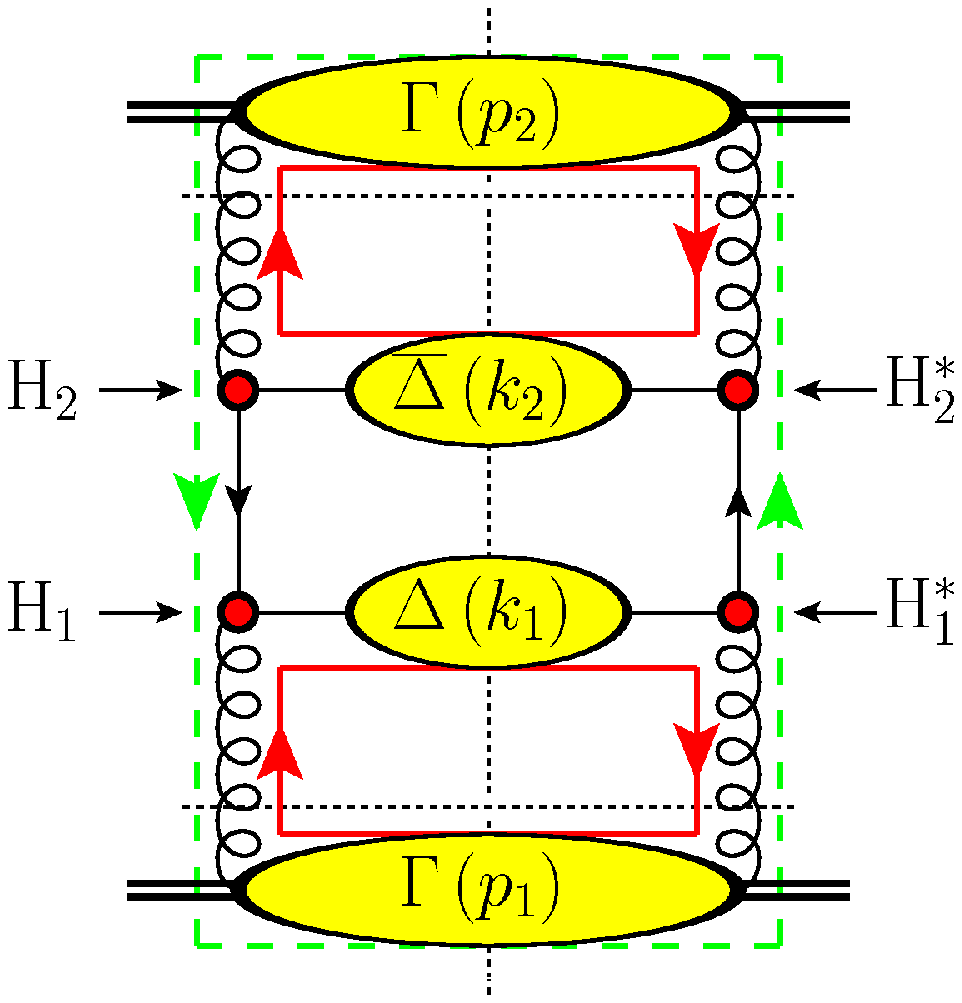,width=0.3\textwidth}
\\[1mm]
(a)\hspace{51mm} (b)
\caption{\label{f:colorflow}
The color flow for (a) $gg\rightarrow \textit{colorless final state}$, e.g. $gg\rightarrow \text{Higgs}$, giving a $U^{[-,-^{\dagger}]}$ gauge link structure. In (b), the leading color flow contribution for this specific $gg\rightarrow q\overline{q}$ diagram can be seen. Other color flow configurations for this diagram are suppressed by factors of $N_c$.}
\end{center}
\end{figure}

For quark correlators there are two types of gauge link structures,
\begin{eqnarray}
\text{type 1:}&&\hspace{5mm}\textrm{Tr}_c \Big\lgroup \psi_i(\xi)\overline \psi_j(0)
\,U_{[0,\xi]}\Big\rgroup, \label{e:qtype1}
\\
\text{type 2:}&&\hspace{5mm}\textrm{Tr}_c \Big\lgroup \psi_i(\xi)\overline \psi_j(0)
\,U_{[0,\xi]}\Big\rgroup\,\frac{1}{N_c}\textrm{Tr}_c \Big\lgroup 
U^{[\text{loop}]}\Big\rgroup. \label{e:qtype2}
\end{eqnarray}
For gluon correlators there are three types of gauge link structures, namely
\begin{eqnarray}
\text{type 1:}&&\hspace{5mm}\textrm{Tr}_c \Big\lgroup F^{n\mu}(0)\,U_{[0,\xi]}^{\phantom{\prime}}\,F^{n\nu}(\xi)\,U_{[\xi,0]}^\prime\Big\rgroup, \label{e:type1} \\
\text{type 2:}&&\hspace{5mm}\textrm{Tr}_c \Big\lgroup F^{n\mu}(0)\,U_{[0,\xi]}^{\phantom{\prime}}\,F^{n\nu}(\xi)\,U_{[\xi,0]}^\prime\Big\rgroup\,\frac{1}{N_c}\textrm{Tr}_c \Big\lgroup U^{[\text{loop}]}\Big\rgroup, \label{e:type2} \\
\text{type 3:}&&\hspace{5mm}\frac{1}{N_c}\textrm{Tr}_c \Big\lgroup F^{n\mu}(0)\,U^{[\text{loop}]}\Big\rgroup\,\textrm{Tr}_c \Big\lgroup F^{n\nu}(\xi)\,U^{[\text{loop}^{\prime}]}\Big\rgroup, \label{e:type3}
\end{eqnarray}
examples of which are given in Fig.~\ref{f:GL}. Which gauge link appears for a particular diagram has been calculated explicitly in Ref.~\refcite{Bomhof:2006dp}, but it is also closely related to the color flow in the process under consideration. When looking at the hard scattering amplitudes, colored particles in the final state typically contribute to gauge links going through plus light-cone infinity, whereas colored particles in the initial state typically contribute to gauge links going through minus light-cone infinity. An example of the latter is the $gg\rightarrow \text{Higgs}$ production through gluon fusion, illustrated in Fig.~\ref{f:FD}a, having a $[-,-^\dagger]$ structure, since the only colored particles are in the initial state. For illustrative purposes, some color flow structures of diagrams in Fig.~\ref{f:FD} have been illustrated in Fig.~\ref{f:colorflow}. We note that in the case of diffractive processes one has operator structures of yet another kind,
\begin{equation}
\text{type 0:}\hspace{5mm}\frac{1}{N_c}\textrm{Tr}_c \Big\lgroup U^{[\text{loop}]}\Big\rgroup, \label{e:type0} \\
\end{equation}
which could play a role in diffractive scattering.\cite{Dominguez:2010xd,Dominguez:2011wm} Traced Wilson loops typically appear whenever a hard scattering diagram contains contributions where it is possible to draw a closed color loop. An example is Fig.~\ref{f:FD}b, which receives a gauge link contribution with a traced Wilson loop $[+,-^\dagger (\square)]$. Note that such diagrams typically contain multiple gauge link contributions, since multiple color flow structures are required in the full description. In Fig.~\ref{f:FD}c the diagram even receives a contribution containing two separate color traces of the form $[+,+^\dagger (\square)(\square^\dagger)]$. The third type of gauge link structures appears whenever there exists a color flow structure which is singlet at the cut, which is e.g. one of the contributions for the diagram in Fig.~\ref{f:FD}d.

The correlators in the Eqs.~(\ref{e:qmatrix}) and (\ref{e:matrix}) including a gauge link can be parametrized in terms of TMD PDFs\cite{Bacchetta:2006tn,Mulders:1995dh} depending on $x$ and $p_\st^2$,
\begin{eqnarray}
&&\Phi^{[U]}(x,p_{\st};n) = \bigg\{
f^{[U]}_{1}(x,p_\st^2)
-f_{1T}^{\perp[U]}(x,p_\st^2)\,
\frac{\epsilon_{\st}^{p_{\st}S_{\st}}}{M}
+g^{[U]}_{1s}(x,p_\st)\gamma_{5}
\nonumber \\&&\mbox{}\qquad
+h^{[U]}_{1T}(x,p_\st^2)\,\gamma_5\,\slashed{S}_{\st}
+h_{1s}^{\perp [U]}(x,p_\st)\,\frac{\gamma_5\,\slashed{p}_{\st}}{M}
+ih_{1}^{\perp [U]}(x,p_\st^2)\,\frac{\slashed{p}_{\st}}{M}
\bigg\}\frac{\slashed{P}}{2},
\label{e:par}
\end{eqnarray}
with the spin vector parametrized as $S^\mu = S_{\sL}P^\mu + S^\mu_{\st} + M^2\,S_{\sL}n^\mu$ and shorthand notations for $g^{[U]}_{1s}$ and $h_{1s}^{\perp [U]}$,
\begin{equation}
g^{[U]}_{1s}(x,p_\st)=S_{\sL} g^{[U]}_{1L}(x,p_{\st}^2)
-\frac{p_{\st}\cdot S_{\st}}{M}g^{[U]}_{1T}(x,p_{\st}^2).
\end{equation}
For quarks, these include not only the functions that survive upon $p_\st$-integration, $f_1^q(x) = q(x)$, $g_1^q(x) = \Delta q(x)$ and $h_1^q(x) = \delta q(x)$, which are the well-known collinear spin-spin densities (involving quark and nucleon spin) but also momentum-spin densities such as the Sivers function $f_{1T}^{\perp q}(x,p_\st^2)$ (unpolarized quarks in a transversely polarized nucleon) and spin-spin-momentum densities such as $g_{1T}(x,p_\st^2)$ (longitudinally polarized quarks in a transversely polarized nucleon). 

For the correlator in Eq.~(\ref{e:matrix}) the expansion in transverse momentum dependent distribution functions (TMDs) at the level of leading twist is given by\cite{Mulders:2000sh,Meissner:2007rx}
\begin{eqnarray}
&&2x\,\Gamma^{\mu\nu [U]}(x{,}p_\st) = 
-g_T^{\mu\nu}\,f_1^{g [U]}(x{,}p_\st^2)
+g_T^{\mu\nu}\frac{\epsilon_T^{p_TS_T}}{M}
\,f_{1T}^{\perp g[U]}(x{,}p_\st^2)
\nonumber\\&&\mbox{}\qquad
+i\epsilon_T^{\mu\nu}\;g_{1s}^{g [U]}(x{,}p_\st)
+\bigg(\frac{p_T^\mu p_T^\nu}{M^2}\,
{-}\,g_T^{\mu\nu}\frac{p_\st^2}{2M^2}\bigg)\;h_1^{\perp g [U]}(x{,}p_\st^2)
\nonumber\\ &&\mbox{}\qquad
-\frac{\epsilon_T^{p_T\{\mu}p_T^{\nu\}}}{2M^2}\;
h_{1s}^{\perp g [U]}(x{,}p_\st)
-\frac{\epsilon_T^{p_T\{\mu}S_T^{\nu\}}
{+}\epsilon_T^{S_T\{\mu}p_T^{\nu\}}}{4M}\;
h_{1T}^{g[U]}(x{,}p_\st^2).
\label{e:GluonCorr}
\end{eqnarray}
In these parametrizations, the process dependence due to the gauge links is absorbed in nonuniversal, gauge link dependent, TMDs.

%%%%%%%%%%%%%%%%%%%%%%%%%%%%%%%%%%%%%%%%%%%%%%%%%%%%%%%%%%%%%%%%%%%%%%%%%%%%%%%
\section{Formalism}
%%%%%%%%%%%%%%%%%%%%%%%%%%%%%%%%%%%%%%%%%%%%%%%%%%%%%%%%%%%%%%%%%%%%%%%%%%%%%%%

For analyzing the process dependent behavior of the TMDs, a formalism for probing these functions has to be used. We will focus on gluons. For analyzing the structure of a single hadron correlator, we use weighting with transverse momenta, defined as
\begin{equation}
\Gamma_{\partial\ldots\partial}^{\alpha_1\ldots\alpha_n[U]}(x)\equiv \int d^2p_{\scriptscriptstyle T}\ p_{\scriptscriptstyle T}^{\alpha_1}\ldots p_{\scriptscriptstyle T}^{\alpha_n}\,\Gamma^{[U]}(x,p_{\scriptscriptstyle T}),
\end{equation}
where we weight the correlator with one or more factors of $p_{\scriptscriptstyle T}$. After a Fourier transformation, these transverse momenta become derivatives in coordinate space, acting on all objects in the correlator depending on the coordinate $\xi_{\scriptscriptstyle T}$, gauge links included.\cite{Buffing:2011mj,Buffing:2012sz} Rather than giving the details for a transverse weighting for multiple factors of $p_{\scriptscriptstyle T}$, we demonstrate the formalism using the single weighted case as an example, which has been described earlier in the Refs.~\refcite{Bomhof:2007xt} and \refcite{Buffing:2013kca},
\begin{equation}
\Gamma_{\partial}^{\alpha [U]}(x)\equiv\int d^2p_{\scriptscriptstyle T}\ p_{\scriptscriptstyle T}^{\alpha}\,\Gamma^{[U]}(x,p_{\scriptscriptstyle T})=\widetilde\Gamma_{\partial}^{\alpha}(x)+C_{G,1}^{[U]}\,\Gamma_{G,1}^{\alpha}(x)+C_{G,2}^{[U]}\,\Gamma_{G,2}^{\alpha}(x).
\label{e:moment1}
\end{equation}
We indicate the partonic operator structure in the correlators with the indices $\partial$ and $G$, the latter of which stands for the gluonic pole contributions and are multiplied with gluonic pole coefficients. These coefficients contain all the process dependence and have been tabulated, see e.g. the Refs.~\refcite{Buffing:2013kca} and \refcite{Bomhof:2006ra}. The basis equations on which the correlators in Eq.~(\ref{e:moment1}) are based are
\begin{eqnarray}
&&\Gamma_{D}^{\mu\nu,\alpha[U]}(x,x-x_1)=\int \frac{d\,\xi{\cdot}P}{2\pi}\frac{d\,\eta{\cdot}P}{2\pi}\ e^{ix_1 (\eta\cdot P)}e^{i (x-x_1)(\xi \cdot P)} \nonumber \\
&&\hspace{23mm} \times\textrm{Tr}\langle P,S|F^{n\mu}(0)\big[U_{[0,\eta]}^{[n]}
iD_{\scriptscriptstyle T}^{\alpha}(\eta)U_{[\eta,0]}^{[n]},U_{[0,\xi]}^{[n]}F^{n\nu}(\xi)U_{[\xi,0]}^{[n]}\big]|P,S\rangle\Big|_{\textrm{\scriptsize{LC}}}, 
\nonumber \\ \label{e:GammaD} \\
&&\Gamma_{F,1}^{\mu\nu,\alpha[U]}(x,x-x_1)=\int \frac{d\,\xi{\cdot}P}{2\pi}\frac{d\,\eta{\cdot}P}{2\pi}\ e^{ix_1 (\eta\cdot P)}e^{i (x-x_1)(\xi \cdot P)} \nonumber \\
&&\hspace{23mm} \times\textrm{Tr}\langle P,S|F^{n\mu}(0)\big[U_{[0,\eta]}^{[n]}
F^{n\alpha}(\eta)U_{[\eta,0]}^{[n]},U_{[0,\xi]}^{[n]}F^{n\nu}(\xi)U_{[\xi,0]}^{[n]}\big]|P,S\rangle\Big|_{\textrm{\scriptsize{LC}}},
\nonumber \\ \label{e:GammaF1} 
\end{eqnarray}
where a third correlator, $\Gamma_{F,2}^{\mu\nu,\alpha[U]}(x,x-x_1)$, similar to $\Gamma_{F,1}^{\mu\nu,\alpha[U]}(x,x-x_1)$ but with an anticommutator of gluonic fields instead rather than a commutator of them, has been omitted. The correlators in Eq.~(\ref{e:moment1}) relate to these basis correlators through the relations
\begin{eqnarray}
&&\Gamma_{D}^{\mu\nu,\alpha}(x)=\int dx_1\ \Gamma_{D}^{\mu\nu,\alpha}(x,x-x_1), \label{e:GammaD1} \\
&&\Gamma_A^{\mu\nu,\alpha}(x)=\int dx_1 \ \textrm{PV}\frac{i}{x_1}\Gamma_{F,1}^{\mu\nu,\alpha}(x,x-x_1), \label{e:GammaA1} \\
&&\widetilde\Gamma_\partial^{\mu\nu,\alpha}(x) \equiv \Gamma_{D}^{\mu\nu,\alpha}(x)-\Gamma_{A}^{\mu\nu,\alpha}(x), \label{e:Gammad1} \\
&&\Gamma_{G,c}^{\mu\nu,\alpha}(x)=\Gamma_{F,c}^{\mu\nu,\alpha}(x,x) \label{e:GammaGc1}.
\end{eqnarray}
The correlators with an index $G$ are the gluonic pole contributions, which correspond to multi-parton correlators with zero momentum gluons, see e.g. the Refs.~\refcite{Efremov:1981sh}, \refcite{Efremov:1984ip}, \refcite{Qiu:1991pp}, \refcite{Qiu:1991wg}, \refcite{Qiu:1998ia} and \refcite{Kanazawa:2000hz}. In this, the index $c$ labels the different color configurations of the partonic operators. The correlator with an index $\partial$ corresponds to a commutator of the partonic operator combination $i\partial =iD -A$ with the gluon field $F(\xi)$.

Transverse weightings can not only be performed at the level of the matrix elements. The expression involving the parametrization of TMDs can be investigated in this way as well, using the definition
\begin{equation}
f_{\ldots}^{g (m)}(x,p_{\scriptscriptstyle T}^2)=\left(\frac{-p_{\scriptscriptstyle T}^2}{2M^2}\right)^m\,f_{\ldots}^{g}(x,p_{\scriptscriptstyle T}^2). \label{e:tm}
\end{equation}
In Eq.~(\ref{e:tm}), the label $m$ represents the number of transverse weightings. See Ref.~\refcite{Buffing:2013kca} for the details of extending this mechanism to higher transverse weightings. Gluonic poles are T-odd, so correlators with an odd number of them are T-odd, while correlators containing an even number of gluonic poles or no gluonic poles at all are T-even.

%%%%%%%%%%%%%%%%%%%%%%%%%%%%%%%%%%%%%%%%%%%%%%%%%%%%%%%%%%%%%%%%%%%%%%%%%%%%%%%
\section{Finding Universal TMDs}
%%%%%%%%%%%%%%%%%%%%%%%%%%%%%%%%%%%%%%%%%%%%%%%%%%%%%%%%%%%%%%%%%%%%%%%%%%%%%%%

By comparing the weighted expressions for operator matrix elements and those of the parametrization, one can identify transverse moments with specific operators. This identification in the previous section can be used to find universal TMD functions. For this we expand the process dependent gluon correlator in terms of definite rank functions depending on $x$ and $p_\st^2$,
\begin{eqnarray}
\Gamma^{[U]}(x,p_{\scriptscriptstyle T}) &\ =\ &
\Gamma(x,p_{\scriptscriptstyle T}^2) 
+ \frac{p_{{\scriptscriptstyle T} i}}{M}\,\widetilde\Gamma_\partial^{i}(x,p_{\scriptscriptstyle T}^2)
+ \frac{p_{{\scriptscriptstyle T} ij}}{M^2}\,\widetilde\Gamma_{\partial\partial}^{ij}(x,p_{\scriptscriptstyle T}^2)
+ \frac{p_{{\scriptscriptstyle T} ijk}}{M^3}\,\widetilde\Gamma_{\partial\partial\partial}^{\,ijk}(x,p_{\scriptscriptstyle T}^2) 
+\ldots 
\nonumber \\ &\quad +&
\sum_c C_{G,c}^{[U]}\bigg(\frac{p_{{\scriptscriptstyle T} i}}{M}\,\Gamma_{G,c}^{i}(x,p_{\scriptscriptstyle T}^2)
+ \frac{p_{{\scriptscriptstyle T} ij}}{M^2}\,\widetilde\Gamma_{\{\partial G\},c}^{\,ij}(x,p_{\scriptscriptstyle T}^2) \nonumber \\
&&\hspace{20mm}+ \frac{p_{{\scriptscriptstyle T} ijk}}{M^3}\,\widetilde\Gamma_{\{\partial\partial G\},c}^{\,ijk}(x,p_{\scriptscriptstyle T}^2) + \ldots\bigg)
\nonumber \\ &\quad +&
\sum_c C_{GG,c}^{[U]}\left\lgroup\frac{p_{{\scriptscriptstyle T} ij}}{M^2}\,\Gamma_{GG,c}^{ij}(x,p_{\scriptscriptstyle T}^2)
+ \frac{p_{{\scriptscriptstyle T} ijk}}{M^3}\,\widetilde\Gamma_{\{\partial GG\},c}^{\,ijk}(x,p_{\scriptscriptstyle T}^2)+ \ldots\right\rgroup
\nonumber \\ &\quad +&
\sum_c C_{GGG,c}^{[U]}\left\lgroup\frac{p_{{\scriptscriptstyle T} ijk}}{M^3}\,\Gamma_{GGG,c}^{ijk}(x,p_{\scriptscriptstyle T}^2)
+ \ldots \right\rgroup + \ldots \, .
\label{e:TMDstructure}
\end{eqnarray}
In this expression the TMD correlators on the rhs are universal objects, multiplied with process dependent gluonic pole factors. Just as for the moments, the functions with an odd number of gluonic poles are T-odd. The rank of the correlators in Eq.~(\ref{e:TMDstructure}) is equal to the rank of the momentum operator in front of it. For each correlator in Eq.~(\ref{e:TMDstructure}) it is therefore clear what its rank and behavior under time-reversal symmetry is.

These two properties can be used to make a classification for the TMDs in the Eqs.~(\ref{e:par}) and (\ref{e:GluonCorr}) as well, where quark TMDs contain contributions up to rank 2 and gluons up to rank 3. In Table~\ref{t:spinhalfcol-1} all correlators are ordered according to their rank and the number of gluonic poles they contain, while in the Tables~\ref{t:quarkPDF-1/2} and \ref{t:gluonPDF-1} the same is done for the quark and gluon TMDs respectively. By comparing these tables an identification can be made, identifying which TMD corresponds to which matrix element. Since it is known how many color structures exist for the matrix elements in Table~\ref{t:spinhalfcol-1}, it is then known how many different universal TMDs exist. An interesting observation is that not only T-odd functions have a process dependence, the T-even functions $h_{1T}^{\perp q}$ for quarks and $h_{1T}^{\perp g}$ for gluons do so as well.

\begin{table}[!tb]
\tbl{The correlators in the expansion in Eq.~(\ref{e:TMDstructure}), ordered by their rank and the number of gluonic poles. Note that the gluonic pole coefficients are equal for correlators in the same row. For quarks the expansion is similar, only there are no rank 3 contributions for quarks and the correlator is indicated with $\Phi$ rather than $\Gamma$.}
{\begin{tabular}{|m{11mm}|m{23mm}|m{23mm}|m{23mm}|m{23mm}|m{23mm}|}
\hline
& \multicolumn{4}{c|}{RANK} \\ \hline
\# GPs &\qquad\quad 0 & \qquad\quad 1 & \qquad\quad 2 & \qquad\quad 3 
\\ \hline
0 
&$\Gamma(x,p_{\scriptscriptstyle T}^2)$
&$\widetilde\Gamma_\partial$
&$\widetilde\Gamma_{\partial\partial}$
&$\widetilde\Gamma_{\partial\partial\partial}$
\\[2pt]
\hline
1
& &$C_{G,c}^{[U]}\Gamma_{G,c}$
&$C_{G,c}^{[U]}\widetilde\Gamma_{\{\partial G\},c}$
&$C_{G,c}^{[U]}\widetilde\Gamma_{\{\partial\partial G\},c}$
\\[2pt]
\hline
2
& & &$C_{GG,c}^{[U]}\Gamma_{GG,c}$
&$C_{GG,c}^{[U]}\widetilde\Gamma_{\{\partial GG\},c}$
\\[2pt]
\hline
3
& & & &$C_{GGG,c}^{[U]}\Gamma_{GGG,c}$
\\[2pt] 
\hline
\end{tabular}
\label{t:spinhalfcol-1}}
\vspace{5mm}\tbl{The assignments of TMD PDFs for quarks as given in Eq.~(\ref{e:par}). The index $c$ for one of the Pretzelocity entries represents the two color configurations that are possible with the same rank and number of gluonic poles.}
{\begin{tabular}{|m{11mm}|p{23mm}|p{23mm}|p{23mm}|p{23mm}|}
\hline
& \multicolumn{4}{c|}{RANK OF TMD PDFs FOR QUARKS} \\ \hline
\# GPs &\qquad\quad 0 & \qquad\quad 1 & \qquad\quad 2 & \qquad\quad 3 
\\ \hline
0 
&$f_1^q$, $g_{1}^q$, $h_{1}^q$
&$g_{1T}^q$, $h_{1L}^{\perp q}$
&$h_{1T}^{\perp q(A)}$
&
\\[2pt]
\hline
1&&$f_{1T}^{\perp q}$, $h_{1}^{\perp q}$&&
\\[2pt]
\hline
2&&&$h_{1T}^{\perp q(Bc)}$&
\\[2pt]
\hline
3&&&&
\\[2pt]\hline
\end{tabular}
\label{t:quarkPDF-1/2}}
\vspace{5mm}\tbl{The operator assignments of the TMD PDFs for gluons that were given in Eq.~(\ref{e:GluonCorr}). The index $c$ for some TMDs indicates the presence of multiple contributions for that TMD as a result of multiple color flow possibilities.}
{\begin{tabular}{|m{11mm}|p{23mm}|p{23mm}|p{23mm}|p{23mm}|}
\hline
& \multicolumn{4}{c|}{RANK OF TMD PDFs FOR GLUONS} \\ \hline
\# GPs &\qquad\quad 0 & \qquad\quad 1 & \qquad\quad 2 & \qquad\quad 3 
\\ \hline
0 
&$f_1^g$, $g_{1}^g$
&$g_{1T}^g$
&$h_1^{\perp g(A)}$
&
\\[2pt]
\hline
1
&&$f_{1T}^{\perp g(Ac)}$, $h_{1T}^{g(Ac)}$
&$h_{1L}^{\perp g(Ac)}$
&$h_{1T}^{\perp g(Ac)}$
\\[2pt]
\hline
2
&&&$h_1^{\perp g(Bc)}$
&
\\[2pt]
\hline
3
&&&&$h_{1T}^{\perp g(Bc)}$
\\[2pt]\hline
\end{tabular}
\label{t:gluonPDF-1}}
\end{table}

One can go one step further, by using Eq.~(\ref{e:TMDstructure}) and finding the gauge link dependent TMDs in terms of the universal ones. This way, one finds for gluons
\begin{eqnarray}
f_{1T}^{\perp g[U]}(x,p_{\scriptscriptstyle T}^2)&=&\sum_{c=1}^2 C_{G,c}^{[U]}\,f_{1T}^{\perp g(Ac)}(x,p_{\scriptscriptstyle T}^2), \label{e:f1Tperp} \\
h_{1T}^{g[U]}(x,p_{\scriptscriptstyle T}^2)&=&\sum_{c=1}^2 C_{G,c}^{[U]}\,h_{1T}^{g(Ac)}(x,p_{\scriptscriptstyle T}^2), \label{e:h1T} \\
h_{1L}^{\perp g[U]}(x,p_{\scriptscriptstyle T}^2)&=&\sum_{c=1}^2 C_{G,c}^{[U]}\,h_{1L}^{\perp g(Ac)}(x,p_{\scriptscriptstyle T}^2), \label{e:h1L} \\
h_1^{\perp g[U]}(x,p_{\scriptscriptstyle T}^2)&=&h_1^{\perp g (A)}(x,p_{\scriptscriptstyle T}^2)+\sum_{c=1}^{4}C_{GG,c}^{[U]}\,h_1^{\perp g (Bc)}(x,p_{\scriptscriptstyle T}^2), \label{e:h1perp} \\
h_{1T}^{\perp g[U]}(x,p_{\scriptscriptstyle T}^2)&=&\sum_{c=1}^2 C_{G,c}^{[U]}\,h_{1T}^{\perp g(Ac)}(x,p_{\scriptscriptstyle T}^2)+\sum_{c=1}^{7}C_{GGG,c}^{[U]}\,h_{1T}^{\perp g(Bc)}(x,p_{\scriptscriptstyle T}^2), \label{e:h1Tperp}
\end{eqnarray}
where the summations indicate the number of color structures that are possible for the different TMDs. In the Eqs.~(\ref{e:f1Tperp})-(\ref{e:h1Tperp}), all functions on the rhs are universal and the gluonic pole factors appearing in these equations have been tabulated in Ref.~\refcite{Buffing:2013kca}.

%%%%%%%%%%%%%%%%%%%%%%%%%%%%%%%%%%%%%%%%%%%%%%%%%%%%%%%%%%%%%%%%%%%%%%%%%%%%%%%
\section{Conclusions}
%%%%%%%%%%%%%%%%%%%%%%%%%%%%%%%%%%%%%%%%%%%%%%%%%%%%%%%%%%%%%%%%%%%%%%%%%%%%%%%

Several of the quark and gluon TMDs as they appear in the expansion of the correlators are process dependent, spoiling universality. We have shown that they can be written as the sum of a finite number of universal TMDs, where the gauge link dependence is isolated in multiplicative gluonic pole factors. For making the identification of which TMDs contain such a process dependence, an expansion of the gluon correlator in terms of transverse moments has been used. By using properties of the elements in this expansion, namely the rank and their behavior under time-reversal symmetry, an identification could be made. Furthermore, we find that also T-even functions can be process dependent. By studying the color flow of hard processes the appropriate gauge link(s) for a given process can be found.\cite{Bomhof:2004aw}

%%%%%%%%%%%%%%%%%%%%%%%%%%%%%%%%%%%%%%%%%%%%%%%%%%%%%%%%%%%%%%%%%%%%%%%%%%%%%%%
\section*{Acknowledgments}
%%%%%%%%%%%%%%%%%%%%%%%%%%%%%%%%%%%%%%%%%%%%%%%%%%%%%%%%%%%%%%%%%%%%%%%%%%%%%%%

This contribution to the conference proceeding combines two separate talks at the QCD Evolution workshop, namely ``TMDs or definite rank for quarks and gluons'' by PJM and ``Universality of gluon TMD distribution functions'' by MGAB. This research is part of the research program of the ``Stichting voor Fundamenteel Onderzoek der Materie (FOM)'', which is financially supported by the ``Nederlandse Organisatie voor Wetenschappelijk Onderzoek (NWO)''. AM thanks the Alexander von Humboldt Fellowship for Experienced Researchers, Germany, for support. Also support of the FP7 EU-programmes HadronPhysics3 (contract no 283286) and QWORK (contract 320389) is acknowledged.

\end{document}